\documentstyle[aps,epsfig]{revtex}
\def \be {\begin{equation}}
\def \ee {\end{equation}}
\def \bea {\begin{eqnarray}}
\def \eea {\end{eqnarray}}
\def \beax {\begin{eqnarray*}}
\def \eeax {\end{eqnarray*}}
\def \a {\hat {a}}
\def \at {\hat {a}^{\dag}}
\def \b {\hat {b}}
\def \bt {\hat {b}^{\dag}}
\def \R {\hat {R}}
\def \A {\hat {A}}
\def \H {\hat {H}}
\def \L {\hat {L}}
\begin{document}
\draft 
\title{Universality class of the restricted solid-on-solid model with hopping}

\author{Su-Chan Park,${}^1$ Jeong-Man Park,${}^2$ and Doochul Kim${}^1$}
\address{${}^1$School of Physics, Seoul National University, Seoul
151-747, Korea\\
${}^2$Department of Physics, 
Catholic University of Korea, Puchon 420-743, Korea
}
\maketitle

\begin{abstract}
We study the restricted solid-on-solid
(RSOS) model with finite hopping distance $l_{0}$, 
using both analytical and numerical methods.
Analytically, we use the hard-core bosonic field theory developed by
the authors [Phys. Rev. E {\bf 62}, 7642 (2000)]
and derive the Villain-Lai-Das Sarma (VLD) equation for 
the $l_{0}=\infty$ case which corresponds to the conserved RSOS
(CRSOS) model and the Kardar-Parisi-Zhang (KPZ) equation for all
finite values of $l_{0}$. Consequently, we find that the CRSOS model
belongs to the VLD universality class and the RSOS models with
any finite hopping distance belong to the KPZ universality class.
There is no phase transition at a certain finite hopping distance
contrary to the previous result. 
We confirm the analytic results using the Monte Carlo
simulations for several values of the finite hopping distance.
\end{abstract}


\section{Introduction}
In recent years, the field of nonequilibrium surface growth has been
investigated using various discrete models and continuous
equations\cite{BS95}. The comprehension of nonequilibrium surface growth 
plays an important role in understanding and controlling many
interesting interface processes, such as vapor deposition\cite{FV85},
crystal growth\cite{PV99}, molecular beam epitaxy (MBE)\cite{HS89}, and so on.
During the MBE growth process, the conserved growth 
condition is applied without defects, such as overhangs and vacancies. Various
discrete conserved models for MBE, describing the kinetic properties of this 
type of surface growth, have been proposed and studied by intensive
numerical simulations.

The main purpose of studying discrete models is to measure
scaling exponents for the kinetic roughening, which determines the asymptotic 
behavior of surface growth on a large length scale in a long time
limit. The important result of the kinetic roughening studies is that a large 
variety of different discrete growth models can be divided into only a few
universality classes. The surface width $W$, which measures the 
root-mean-square fluctuation of the surface height,
scales as 
\be
W(L,t) \sim L^\alpha f(t/L^z),
\ee
where the asymptotic behavior of the scaling function $f(x)$ is constant
for $x\gg 1$ and $x^\beta$ for $x\ll 1$ with $\beta = \alpha / z$. The 
scaling behavior of the growth is characterized by three exponents:
the roughness exponent $\alpha$, the growth exponent $\beta$, and the 
dynamical exponent $z$. These exponents determine the universality
class.

In the coarse-grained picture, evolution of the growing surface is usually
described by a stochastic differential equation (SDE) for the height
variable $h({\bf x},t)$ as a function of the surface
coordinate and time. For discrete models of MBE growth, several SDEs were
suggested and it was generally believed that there is a correspondence
between discrete growth models and continuum SDEs.
The common way of establishing the link
between discrete models and continuous equations is a simple
comparison of critical exponents determined from computer simulations of
the discrete model with exponents for the continuous equation.

There have also been attempts to establish the correspondence 
in an explicit way.
The systematic procedure for establishing a continuous
equation corresponding to discrete models, starting from
the master equation in discrete space was proposed by 
Vvedensky {\it et al.}\cite{VZLW93}
and has been successfully applied to the derivation of growth equations
for some discrete models, including the solid-on-solid (SOS) model, the 
restricted solid-on-solid (RSOS) model, as well as the 
Wolf-Villain and Das Sarma-Tamborenea models\cite{PK95,PK96,HG98}.
However, there are several difficulties with
this procedure; in particular, in converting from the
equation system for a discrete set of heights to an equation for
a continuous function $h({\bf x},t)$, the procedure requires the 
regularization step, in which the nonanalytic quantities are expanded and
replaced with analytic quantities, i.e., the step function is approximated
by an analytic shifted hyperbolic tangent function expanded in a Taylor
series. However, the form of the regularized function is uncertain, 
and different choices of this function lead to different results.
Some forms of the regularized function have been suggested, but the 
problem of a proper choice of a regularization scheme has not been
discussed.

To overcome this kind of uncertainty, the authors proposed a new schematic
formalism\cite{PKP00}, 
deriving the continuous equations, such as the Edwards-Wlikinson 
(EW)\cite{EW82} and 
Kardar-Parisi-Zhang (KPZ)\cite{KPZ86}
equations, from the body centered solid-on-solid model and the RSOS model.
In this paper, we apply our formalism to a new kind of MBE growth
model proposed by Kim, Park, and Kim\cite{KPK94}. 
This model allows the deposited
particle to relax to the nearest site where the RSOS condition on neighboring
heights is satisfied and has the conserved growth condition constraint, 
which means the deposited particles are possible to hop for
an infinite distance until they eventually find the site with the RSOS
condition satisfied. 
Applying our formalism to the above conserved RSOS (CRSOS) model, 
we not only derive the Villain-Lai-Das Sarma (VLD)
equation\cite{VLD91} for the model 
which belongs to a different universality class from the EW and
KPZ equations, but also we are able to predict the coefficients in the VLD
equation, which was not possible by other methods.

Observing that the RSOS model belongs to the KPZ class and the CRSOS model
belongs to the VLD class, we went one step further to study the RSOS
model with the finite range hopping (RSOS/H). In this RSOS/H model, it
is possible for the deposited particles to hop a finite distance $l_0$ until 
they find the site with the RSOS condition satisfied. If they fail to
find the site with the RSOS condition satisfied within the distance
$l_0$  in both directions, the deposition process is
rejected. The RSOS model corresponds to the RSOS/H model with 
$l_0=0$ and to the CRSOS model with $l_0 = \infty$. We apply our formalism
to the RSOS/H model with $l_0$ finite and find that this model 
belongs to the KPZ class, contrary to the previous report 
by Kim and Yook\cite{KY97},
who concluded that there is a phase transition between the KPZ class to the 
VLD class along the parameter $l_0$.

In Sec. \ref{an} our formalism to derive the continuous
equation from the discrete model is briefly explained and the 
procedure of derivation is described. The detailed calculations are attached
in the appendices. Extensive numerical simulations are presented in 
Sec. \ref{num} and the summary and discussion are given in Sec. \ref{dis}.
\section{Derivation of the stochastic equation}
\label{an}
In this section, we derive continuous 
equations for the one-dimensional RSOS/H model with a hopping
distance $l_0$ and for the CRSOS model corresponding to $l_0 = \infty$.
We restrict ourselves to the case wherein the height difference
between two nearest neighbors is not larger than 1.
For a succinct description of the dynamics, we introduce the nomenclature
that if a site $m$ satisfies the condition $|h_m+1 -h_{m\pm 1}|\le 1$,
this site is called ``stable.''  Following this nomenclature,
the growth algorithm of the one-dimensional RSOS/H model
is as follows: 
(i) We choose a site $m$ randomly.
(ii) Sites from $m-l_0$ to $m+l_0$ are examined to determine if they
are stable sites.
(iii) If a stable site is found within the interval from $m-l_0$ to $m+l_0$,
a new particle is deposited to the nearest stable 
site from $m$ ($m$ itself can be a candidate for deposition).
However, if stable sites are nonexistent in the examined interval,
the particle drop is rejected and the system remains unchanged.
After this try, the time is increased by $1/L$, where $L$
is the system size.
We assume periodic boundary conditions.

Since the height difference between two nearest neighbors 
is restricted not to be larger than 1, 
the RSOS/H model is mapped onto the reaction-diffusion system
of hard-core particles with two species.
The step-up (-down) is mapped to an $A$ ($B$) particle.
If two nearest neighbor sites have equal height, a
particle vacuum is located between these two sites.
The site
where the particles reside is labeled by an integer, and
the site for height by a half-integer. This mapping is depicted in 
Fig.\ref{config}.
According to this mapping, the dynamics of the RSOS/H model
can be described by the (imaginary time)
Sch\"ordinger equation 
$(\partial / \partial t) | \Psi;t\rangle = -\H | \Psi ; t \rangle$
for the state vector $|\Psi ; t \rangle \equiv \sum_C P(C;t) |C \rangle$,
where $P(C;t)$ is the probability with which the system is in state 
$C$ at time $t$, and 
$\H$, called a Hamiltonian, is an evolution operator
\bea
\H = \sum_n \left ( \hat I - \R_n - \A_n \right ) \L_n ,
\label{hamiltonian}
\eea
where
\bea
\L_n = \hat I + {1\over 2} \sum_{j=1}^{2 l_0} \left (
\prod_{k=1}^j \R_{n+k}+ \prod_{k=1}^j \R_{n-k} \right ),
\eea
and 
\be
\begin{array}{ccl}
\R_n& =& \at_n \a_n + \bt_{n+1} \b_{n+1} - \at_n \a_n \bt_{n+1} \b_{n+1},
\vspace{0.cm}\\
\\
\A_n& =& \left ( \at_n + \b_n \right ) \left ( \a_{n+1} + \bt_{n+1} \right ).
\end{array}
\ee
The role of the rejection operator
$\R_n$ is to check the stability of site $n+{1\over2}$ 
, that is, if a configuration $|C\rangle$ has a stable  site at $n+{1\over2}$, 
$\R_n | C \rangle = 0$ and otherwise
$\R_n | C \rangle =|C \rangle$.
The adsorption operator 
$\A_n$ describes the configuration change
after a successful deposition.
$\a_n$ ($\b_n$) is the annihilation operator of an $A$ ($B$) particle
at site $n$ and $\at_n$ ($\bt_n$) is the corresponding creation operator,
satisfying the mixed commutation relations presented in Ref.\cite{PKP00},
\bea
\a_n \at_n = \b_n \bt_n = \hat I - \at_n \a_n - \bt_n \b_n,\quad
\a_n \b_n = \at_n \bt_n = 0.
\eea
All operators at different sites commute with each other.

To find the SDE for the 
RSOS/H model, we apply the method recently 
introduced by the authors\cite{PKP00}.
First we assume the existence of the lattice version of the SDEs 
in terms of density.
Those equations are supposed to take the forms
\bea
\partial_t a_n&=& C_n^A\left ( \{ a\},\{b\}\right) + 
\sum_m\left [ g_{nm}^{AA} \xi_m^A(t) + g_{nm}^{AB}\xi_m^B (t)\right ],
\label{SDEa}\\
\partial_t b_n&=& C_n^B\left (\{a\},\{b\}\right) +
\sum_m \left [ g_{nm}^{BA} \xi_m^A(t) + g_{nm}^{BB}\xi_m^B (t)\right ],
\label{SDEb}
\eea
where $\partial_t \equiv \partial / \partial t$ and 
$\xi_n^A$,  $\xi_n^B$ are white noises with correlation
\bea 
&\langle \xi_n^X(t) \xi_m^{X'}(t') \rangle = \delta_{nm}
\delta_{X,X'} \delta(t-t')& \nonumber\\
&(X,X' = {\rm either~}A~{\rm or}~B),&
\eea
where $\delta_{nm}$ and $\delta_{X,X'}$ are Kronecker deltas and 
$\delta(t-t')$ is the Dirac delta function.
The matrix $g$ is related to the Kramers-Moyal coefficient in such a way 
that 
\bea
&\sum_{r,X''} g_{nr}^{XX''} g_{mr}^{X'X''} =
C_{nm}^{XX'}\left ( \{a\},\{b\} \right )&
\nonumber\\
&(X,X',X'' = {\rm either~}A~{\rm or}~B).&
\eea
Here we are using the It\^o interpretation.
The field $a$ ($b$) in the curly bracket represents
the density of species $A$ ($B$) at all sites.
From here on, without a hat above itself a mathematical
symbol is a pure number as opposed to an operator.
$a_n$ should not be confused with $\a_n$. The former is a density at 
site $n$ that runs over real numbers, while the latter is an annihilation
operator.
By requiring that the noise average of observables in Eqs. (\ref{SDEa})
and (\ref{SDEb})
has the same behavior with the ensemble average of the number operator,
we find\cite{PKP00}
\bea
\langle C_n^A  \rangle_t &=&
\langle [H, \at_n \a_n] \rangle_t,\nonumber\\
\langle C_n^B  \rangle_t &=&
\langle [H, \bt_n \b_n] \rangle_t,\nonumber\\
\langle C_{nm}^{AA}
\rangle_t &=& \left \langle \left [\at_n \a_n, [H, \at_m \a_m] \right ]
\right \rangle_t,
\label{commutation}\\
\langle C_{nm}^{AB}
\rangle_t &=& \langle C_{mn}^{BA} \rangle_t
= \left \langle \left [\at_n \a_n, [H, \bt_m \b_m]\right ] \right \rangle_t
,\nonumber
\\
\langle C_{nm}^{BB}
\rangle_t &=& \left \langle \left [\bt_n \b_n, [H, \bt_m \b_m] 
\right ] \right \rangle_t,
\nonumber
\eea
where the $\langle\cdots \rangle_t$ on the left hand side represents the 
average over noise at time $t$ 
and that on the right hand side stands for the ensemble average.
The arguments of the Kramers-Moyal coefficients are dropped for brevity.

As presented in Ref.\cite{PKP00}, the  ensemble average of any operator
can be interpreted as an average of number operators due to 
the property of the projection state $\langle \cdot |$, which is defined
as a sum over all possible microscopic configurations and is itself
a left eigenstate of $\H$ with eigenvalue $0$,
and in turn, the ensemble average of a number operator is mapped 
to the noise average of density. This procedure leads us 
to find the Kramers-Moyal coefficients $C_n^X$, $C_{nm}^{XY}$,
in terms of the density fields. We call this procedure
{\it figurization}, which means ``expression in number.''
To represent the {\it figurization}, we use the symbol $\longmapsto$
to the left of which is an operator (or a product of operators) and 
to the right of which is the corresponding density representation. 

To complete the derivation, we perform the commutation relations
between the Hamiltonian and the density operators and so forth:
\bea
\left [ \H , \at_n \a_n \right ] &=& \at_n \left ( \a_{n+1} + \bt_{n+1} \right )
\L_n 
- \a_n \left ( \at_{n-1} + \b_{n-1} \right ) \L_{n-1} ,
\label{den_commute_a}
\\
\left [ \H, \bt_n \b_n \right ] &=& -\b_n \left ( \a_{n+1} +\bt_{n+1} 
\right )\L_n
 + \bt_n \left ( \at_{n-1} + \b_{n-1} \right ) \L_{n-1},
\label{den_commute_b}\\
\left [ \at_l \a_l, \left [ \H , \at_n \a_n \right ] \right ] &=& 
\delta_{ln} \left ( \at_n \left ( \a_{n+1} + \bt_{n+1} \right )
\L_n 
+ \a_n \left ( \at_{n-1} + \b_{n-1} \right ) \L_{n-1}\right ) \nonumber\\
&&-\delta_{l,n+1} \at_n\a_{n+1}\L_n - \delta_{l,n-1} 
\a_n  \at_{n-1} \L_{n-1},\\
\left [ \bt_l \b_l,  \left [ \H, \bt_n \b_n \right ] \right ] &=&
\delta_{ln} \left ( \b_n \left ( \a_{n+1} +\bt_{n+1} 
\right )\L_n
 + \bt_n \left ( \at_{n-1} + \b_{n-1} \right ) \L_{n-1} \right ) \nonumber\\
&& - \delta_{l,n+1} \b_n \bt_{n+1} \L_n
- \delta_{l,n-1} \bt_n \b_{n-1} \L_{n-1},\\
\left [ \at_l \a_l, \left [ \H, \bt_n \b_n \right ] \right ] &=&
\delta_{l,n+1} \b_n  \a_{n+1} \L_n
+ \delta_{l,n-1} \bt_n  \at_{n-1}  \L_{n-1},
\eea
where $\delta_{nl}$ is a Kronecker delta.
Following the {\it figurization}, we find the Kramers-Moyal coefficients.
For a later purpose, we give some examples of the {\it figurization}.
The {\it figurization} of $\R_n$
is $\R_n \longmapsto R_n\equiv a_n + b_{n+1} - a_n b_{n+1}$,
and the symbolic representation of the {\it figurization} 
of the product of $\R$'s is
\bea
\prod_{k=1}^j \R_{n+k} \longmapsto R_n^j,
\eea
where the superscript $j$ should not be confused with the power.
When $j=1$, $R_n^1$ is denoted as $R_{n+1}$. 
$R_n^j$ are given by the following recursion relations:
\bea
R_n^j &=& a_{n+1} R_{n+1}^{j-1} + (1 - a_{n+1} ) \prod_{k=2}^{j+1} b_{n+k},
\label{recursive1}\\
R_{n-j-2}^j &=& b_{n-1} R_{n-j-2}^{j-1} + (1 - b_{n-1}) \prod_{k=2}^{j+1}
 a_{n-k},
\label{recursive2}
\eea
where $j \ge 1$ and we define $R_n^0 \equiv 1$.
The physical meaning of Eqs. (\ref{recursive1}) and 
(\ref{recursive2}) is as follows:
We divide the situation that prohibits the height increase at sites $
n+{3 \over 2}, n+{5\over 2}, \cdots, n+j+{1\over 2}$
by the condition at site $n+1$. If
there is an $A$ particle at site $n+1$, the height increase is suppressed
at site $n+{3 \over 2}$ irrespective of the condition at site $n+2$. Hence the
first term of Eq. (\ref{recursive1}) follows. If there is no $A$ particle
at site $n+1$,
to suppress the deposition at site $n+{3 \over 2}$, 
there must exist a $B$ particle
at site $n+2$ and this should be continued until the site $n+j+1$, because
at site $n+k$ ($2 \le k \le j$) no $A$ particle is present; this
condition is represented by the second term.
To comprehend Eq. (\ref{recursive2}), we only have to perform the
mirror transformation relative to site $n$.
By the mirror transformation relative to  a site $n$, we mean 
the exchange of $a$ and $b$ ($a \leftrightarrow b$), followed by 
$n+k \leftrightarrow n-k$. Under this transformation,
$R_{n+k}^j$ changes into $R_{n-k-j-2}^j$ for an arbitrary $k$.
The mirror transformation of Eq. (\ref{recursive1}) 
is Eq. (\ref{recursive2}).
 
With these notions, we will find the SDE of the RSOS/H model.
At first, the deterministic part of SDE is found.
Since the main concern is not the respective dynamics of 
the $A$ and $B$ particles,
but it is $D_n \equiv a_n - b_n$ (the local slope) and $S_n = a_n + b_n$ 
(the slope density), we will write the SDE for $D$ and $S$ rather
than for the $A$, $B$ particles.
The Kramers-Moyal coefficient $C_n^D$
is obtained by subtracting
Eq. (\ref{den_commute_b}) from Eq. 
(\ref{den_commute_a}) and $C_n^S$ 
by adding Eq. (\ref{den_commute_b}) and (\ref{den_commute_a}) 
followed by the {\it figurization}.
\bea
C_n^D &=& {1\over 2}\sum_{j = 1}^{2 l_0} \left [
R_{n-2}^j - 2 R_{n-1}^j + R_n^j 
- \left (R_{n-j-2}^j - 2 R_{n-j-1}^j + R_{n-j}^j \right ) \right ]
\nonumber
\\
&+&{ 1\over 2} \left ( R_{n-2}^{2l_0 +1} - R_{n-1}^{2l_0 + 1} 
+ R_{n-2l_0 - 2}^{2l_0 + 1} - R_{n-2l_0 - 1}^{2l_0 +1} \right ),
\label{KM_of_height}
\\
C_n^S &=& 2 - 3 (a_n + b_n)
- ( b_n a_{n+1} + a_n b_{n-1} - 3 b_n b_{n+1} - 3 a_n 
a_{n-1} )\nonumber\\
&+&{1\over 2} \sum_{j=1}^{2l_0} \left ( 
R_n^j - R_{n-2}^j - R_{n-j}^j + R_{n-j-2}^j \right )
- b_n ( 1 - S_{n+1} ) b_{n+2} 
- a_n ( 1 - S_{n-1} )a_{n-2}\nonumber \\
&-& {1\over 2} ( R_{n-1}^{2 l_0 + 1}+
R_{n-2 l_0-1}^{2 l_0 + 1}+R_{n-2}^{2 l_0 + 1}+R_{n-2 l_0 -2}^{2 l_0 + 1})
\nonumber\\
&-&\sum_{j=1}^{2l_0 - 1} \left [
b_n a_{n+1} R_{n+1}^j + b_n ( 1 - S_{n+1} ) \prod_{k=2}^{j+2} b_{n+k}
+ b_n ( 1 - b_{n+1} ) R_{n-j-2}^j \right ]
\nonumber\\
&-&\sum_{j=1}^{2l_0 - 1} \left [
a_n b_{n-1} R_{n-j-3}^j + a_n ( 1 - S_{n-1} ) \prod_{k=2}^{j+2} a_{n-k}
+ a_n ( 1 - a_{n-1} ) R_n^j \right ].
\label{KM_of_step}
\eea
As pointed out in Ref.\cite{NdN97}, the mass term in Eq. 
(\ref{KM_of_step}) makes the step density saturate fast.
As a result, the step density $S$ becomes a slave field of
the slope $D$ and takes the form
\be
S = 2 \rho (l_0) + \mu_0 (l_0) \partial D + \theta_0 (l_0) D^2 + \cdots,
\label{step_density}
\ee
where $2 \rho (l_0)$
is the (mean-field) stationary step density and $\mu_0$ and $\theta_0$
may depend on $l_0$. 
Here $\partial \equiv (\partial / \partial x)$.
Since this system has the
mirror symmetry whose continuum version is invariant
under the transformations $D \rightarrow -D$ and $\partial \rightarrow
-\partial$, we do not expect the occurrence of
$D$ in Eq. (\ref{step_density}). 
The parameters $\rho (l_0)$, $~\mu_0 (l_0)$, and $\theta_0 (l_0)$ are 
determined by the stationarity of Eq. (\ref{KM_of_step}).
In Appendix \ref{Rj}, we obtain $\rho$, $~\theta_0$, and $\mu_0$ as 
functions of $l_0$ and show the approximate solutions of these parameters.
We can now rewrite Eq. (\ref{KM_of_height}) in terms of the height
field $h$ using Eq. (\ref{step_density}) and
\be
D = \partial h.
\label{slope}
\ee
The last task is to determine the noise strength.
This is accomplished using Eqs. (\ref{SDEa}) and (\ref{SDEb}).
After eliminating $S$ in favor of $D$, there is only one kind of
noise with strength
\be
C_{nm} = C_{nm}^{AA} + C_{nm}^{BB} - C_{nm}^{AB} - C_{nm}^{BA}.
\label{noise_strength}
\ee 
Using {\it figurization} of Eq. (\ref{commutation}) and keeping the 
most relevant term, we find
\begin{equation}
C_{nm} = - ( 1 - \rho )^2 L 
( \delta_{n,m+1} - 2 \delta_{nm} + \delta_{n,m-1}) + \cdots,
\end{equation}
where 
\begin{equation}
L = \sum_{l=0}^{2 l_0} \rho^l( 1 + l - \rho l).
\end{equation} 

For finite $l_0$ we found 
\bea
\partial_t h = v_\infty + \nu \partial^2 h + {\lambda \over 2} (\partial h)^2
+ \xi(x,t),\label{KPZeq}\\
\langle \xi(x,t) \xi(x',t') \rangle = D_{\xi \xi} \delta(x-x') \delta(t-t'),
\nonumber
\eea
where
\bea
v_\infty &=&1-R(2 l_0 + 1) \simeq1-\rho_\infty^{2 l_0} \left [
{ 1 \over 2} + \left (\sqrt{2} - 1 \right ) l_0 \right ] + 
O\left (\rho_\infty^{4 l_0}\right ),
\nonumber
\\
\nu &=&-\mu(2 l_0 +1)\simeq \rho_\infty^{2 l_0} (l_0 + 1 ) (2 l_0 + 1 )
\left ( {\sqrt{2} \over 6}l_0 + { 3 -\sqrt{2} \over 4} \right ) + 
O\left (\rho_\infty^{4 l_0}\right ),
\\
\lambda &=&-2 \theta(2 l_0 +1)\simeq
- \rho_\infty^{2 l_0} (1 + l_0)( 1 + 2 l_0)
\left (1 + {\sqrt{2} + 1 \over 3} l_0 \right )+
O\left (\rho_\infty^{4 l_0}\right ),
\nonumber\\
D_{\xi\xi}&\simeq& { 2 \sqrt{2} - 1\over 2} \left [ 1 - \rho_\infty^{2 l_0} \left (
{6 + 5 \sqrt{2} \over 28} + {19 \sqrt{2} - 22 \over 14} l_0 \right ) \right ]
+
O\left (\rho_\infty^{4 l_0}\right ),
\nonumber
\eea
with $\rho_\infty = (2 -\sqrt{2})/2$.
The numerical values of $\rho(l_0)$, $~\nu(l_0)$, and $\lambda(l_0)$ are given
for several values of $l_0$ in Table I.
For infinite $l_0$
\bea
\partial_t h = \tilde v_\infty 
- \tilde \nu \partial^4 h + \tilde \lambda \partial^2 
(\partial h)^2 + \tilde \xi(x,t),\label{VLDeq}\\
\langle \tilde \xi(x,t) \tilde \xi(x',t') \rangle = \tilde D_{\xi\xi}
\delta(x-x') \delta (t-t'),\nonumber
\eea
where
\bea
\tilde v_\infty &=& 1,\nonumber\\
\tilde \nu &=& {21 - 12 \sqrt{2}\over 2},\nonumber\\
\tilde \lambda &=& {10 - 3 \sqrt{2} \over 2},
\\
\tilde D_{\xi\xi} &=& { 2 \sqrt{2} - 1 \over 2}.
\nonumber
\eea

Equation (\ref{KPZeq}) directly shows that the RSOS/H model, for 
any finite $l_0$, belongs
to the KPZ class, and Eq. (\ref{VLDeq}) suggests that the CRSOS model
is described by the VLD equation. 
However, the first line of Eq. (\ref{KM_of_height}) 
has the form
\be
\sum_{j = 1}^{2 l_0}  [
R_{n-2}^j - 2 R_{n-1}^j + R_n^j 
-  \{{\rm mirror~terms~of}~  (R_{n-2}^j - 2 R_{n-1}^j + R_n^j 
) \}  ].
\label{Firstline}
\ee
which has the following implications:
If the continuum version of $R_{n-2}^j - 2 R_{n-1}^j + R_n^j$ has 
a nonvanishing coefficient of $\partial^2 (D^{2 r +1})$ with a non-negative
integer $r$, 
it is not certain that Eq. (\ref{Firstline}) is a lattice 
version of $\partial^3$; consider the mirror transformation in the 
continuum limit.
The occurrence of $\partial^2 (D^{2 r +1})$ in Eq. (\ref{Firstline})
is directly related to the appearance of $\partial (\partial h)^{2r+1}$
in Eq. (\ref{VLDeq}), which generates an EW term 
by the dynamic renormalization group\cite{KG96}.
Appendix \ref{Proof_No_KPZ} shows that this
is not the case.
The vanishing of $D^{2 r +1}$ in the continuum limit of
$R_{n+k}^j$ guarantees the vanishing of $\partial^2 (D^{2 r +1})$ 
in $R_{n-2}^j - 2 R_{n-1}^j + R_n^j$.
In view of this, we conclude that the continuum equation of the
CRSOS model is the VLD equation.

To confirm this conclusion, we have performed Monte Carlo
simulations as outlined  in Sec. \ref{num}.
\section{Numerical study}
\label{num}
Although the RSOS/H model was studied numerically 
by Kim and Yook\cite{KY97}, their
results are contradictory to our derivation. As a result, we need
to perform extensive numerical simulations to verify our results.
In the derivation, we found that the coefficients of the EW terms
and the KPZ terms are vanishingly small, though finite, for large $l_0$.
Hence, we may find a crossover of roughness exponents
from VLD ($\alpha_{\rm  vld} \simeq 0.95$) to
KPZ ($\alpha_{\rm  kpz} = 0.5$).
We are preoccupied with the numerical observation of this crossover.

In Fig. \ref{alpha}, we draw the saturated width $W_{\rm sat}$
as function of the system size $L$ for some $l_0$'s. 
The system sizes are 64, 90, 128, 180, 256, 360, 512, 720, and 1024.
For relatively small system sizes, the roughness exponents are 
near to the values reported in Ref. \cite{KY97}. As expected from our 
derivation, we see a crossover for large system sizes. 

To clarify the crossover behavior, a scaling plot is given
in Fig. \ref{1dcross}.
The anticipated scaling form of the saturated width is
\be
W_{\rm sat} ( l_0 ,L ) = L^{\alpha_{\rm  vld}} g(l_0^\gamma / L),
\ee
where $\gamma$ is the crossover exponent.
The asymptotic behavior of the scaling function $g$ is expected to be
\be
g(x) \sim 
\left \{
\begin{array}{lccccc}
{\rm const}&~~&{\rm when} &x&\rightarrow& \infty\\
x^{\alpha_{\rm  vld} - \alpha_{\rm  kpz}}&~~&{\rm when}& x& \rightarrow& 0
\end{array}.
\right .
\ee
The best fit for the data set shown in Fig. \ref{alpha} corresponds to
$\alpha_{\rm  vld} = 0.9$ and $\gamma = 2.0$. The fitting parameter
$\alpha_{\rm  vld}$ obtained is somewhat smaller than the known value of
the roughness exponent of the CRSOS model. This is most likely
due to the smallness of $l_0$. For example, the data
for $l_0 = 10$ in Fig. \ref{alpha} yield 0.86 rather than 0.95.
The meaning of $\gamma$ is as follows:
When the system size is sufficiently large, we expect 
the width of the characteristic mountain to be
$\sim L^{\alpha_{\rm  kpz}}$. 
Due to the RSOS condition, the linear size of the mountain is also
expected to be $\sim L^{\alpha_{\rm  kpz}}$. 
The smaller the system size, the less the rejection events occur
due to the shrinking of the characteristic mountain.
If $l_0$ is comparable with the linear size of the 
characteristic mountain ($L^{\alpha_{\rm kpz}}$), the system starts to
behave differently. Eventually, the system with small size, 
$L^{\alpha_{\rm kpz}} \ll l_0$, cannot feel the existence of $l_0$.
Thus the crossover length $L^*$ is expected to be
$l_0^{1/{\alpha_{\rm kpz}}}$, that is, $\gamma = 1/{\alpha_{\rm kpz}} = 2$.
\section{Summary and Discussion }
\label{dis}
We studied the RSOS/H model using both analytical and numerical methods.
We derived the continuum equation
for the microscopic discrete model analytically, and 
found coefficients of the EW term $\partial^2 h$, 
the KPZ term $(\partial h)^2$,
as well as the VLD term $\partial^2 (\partial h)^2$.
We observed that the coefficients of the EW and KPZ terms
behave as $\sim a_0^{2 l_0} l_0^3$ for sufficiently large $l_0$, 
which is consistent with the previous numerical study\cite{KY97}.
Accordingly, we concluded that the RSOS/H model for any finite $l_0$ 
eventually belongs to the KPZ class and the CRSOS model belongs to the VLD
class. Numerically, we reported the crossover from 
the VLD class to the KPZ one, which confirms our derivation. 
Moreover, we found a crossover exponent $\gamma$, which is argued to be 2.

Besides these studies, we can offer
an (nonrigorous plausible) argument to anticipate the universality
class of the RSOS/H model by employing 
the block spin concept of Kadanoff's. Consider a system
with linear size $L$ and hopping distance $l_0$.
Similar to the block spin in the Ising model, we introduce a coarse-graining
parameter $b$ which blocks the $b$ sites by one. Although
the exact transformation of coarse-graining cannot be determined, we
expect that if it exists, $l_0$ may be renormalized as $\sim l_0/b^{\gamma '}$.
Hence, we expect that the stable fixed point corresponds to $l_0=0$, which 
is the KK model\cite{KK89} and the unstable fixed point corresponds
to $l_0 = \infty$,
which is the CRSOS model\cite{KD94}.

\section*{Acknowledgments}
We are grateful to Byungnam Kahng for helpful discussions.
This work was supported by Grant No.2000-2-11200-002-3 from 
the BRP program of the KOSEF. 

\appendix
\section{~Continuum limit and determination of $\theta_0$ and $\mu_0$}
\label{Rj}
In this appendix, we derive the continuum limit of $R_{n + k}^l$ 
where $n$ is the reference point. One may obtain the continuum limit
directly from Eq. (\ref{recursive1}), but we do not follow this route.
Rather, we make recursion relation about the parameters
that appear in front of the field (see below). 
The use of Eq. (\ref{step_density}) makes it possible to represent
the continuum limit in terms of $D$, which is directly related to the
height by Eq. (\ref{slope}). 
We can find the correct coefficient that appears
in Eq. (\ref{step_density}) after we find the continuum limit
of $R_{n+k}^l$. 

The continuum limit of $R_{n+k}^l$ takes the form
\be
R_{n+k}^l \doteq R(l) + \gamma (l,k) D + \theta (l,k) D^2 + \mu(l,k)
 \partial D + \cdots,
\label{cont_lim}
\ee
where the implicit dependence on $l_0$ is assumed and the argument of D
is dropped for simplicity.
The symbol ``$\doteq$'' represents the continuum limit of a quantity. 
Equation (\ref{recursive1}) allows us to find the recursion relations.
Before going further, the explicit form of $R(l)$ in the case
of no tilt boundary condition is found.
The recursion relation becomes
\bea
R(l) &=& \rho R(l-1) + ( 1 - \rho ) \rho^l,\nonumber\\
\Rightarrow \rho^{-l} R ( l ) &=& \rho^{-(l-1)} R(l-1) + 1 - \rho
\eea
with an (sort of) initial condition $R(1) = 2 \rho - \rho^2$. $\rho$ is the
abbreviation of $\rho(l_0)$ in Eq. (\ref{step_density}).
It is trivial to find the solution that reads
\be
R(l) = \rho^l + \rho^l ( 1 - \rho ) l.
\ee
To find the recursion relations, we need a continuum limit of $P_l^k$, which
is defined as
\be
P_l^k \equiv \left ( 1 - a_{n+k+1} \right )
\prod_{m=2}^{l+1} b_{n + k + m} = 
P_{l-1}^k b_{n+k+l+1}.
\label{recursive_of_P}
\ee
One may directly calculate the continuum limit of $P_l^k$, but we
follow another path. Let the continuum limit of $P_l^k$ be
\be
P_l^k \doteq  (1-\rho)\rho^l + \gamma_1 (l,k) D + \theta_1 (l,k) D^2
+\mu_1 (l,k) \partial D +\cdots. 
\label{cont_P}
\ee
From Eq. (\ref{recursive_of_P}), we find the recursion relations ($l\ge 2$)
\bea
{ \gamma_1  (l,k)\over \rho^l} &=&
{\gamma_1(l-1,k)\over \rho^{l-1}} - {1 - \rho \over 2\rho},
\nonumber\\
{\theta_1 (l,k)\over \rho^l} &=& 
{\theta_1(l-1,k)\over \rho^{l-1}} - 
{\gamma_1(l-1,k)  \over 2 \rho^l} + \theta_0 { 1 - \rho \over 2\rho},
\label{relation_P}\\
{\mu_1 (l,k)\over \rho^l}  &=& {\mu_1 (l-1,k)\over \rho^{l-1}} 
+ ( \mu_0 - k-l-1 ) { 1 - \rho \over 2\rho}, 
\nonumber
\eea
with initial conditions
\bea
\gamma_1(1,k) &=& - { 1 \over 2},\nonumber\\
 \theta_1 (1,k) &=&  { 1 \over 4} ( 1 + 2 \theta_0 - 4 \rho \theta_0 ),
\label{initial_P}\\
\mu_1 (1,k) &=& { 1\over 2} ( \mu_0 -k-2- 2\rho \mu_0 + \rho ),
\nonumber
\eea
$\theta_0$ and $\mu_0$ are functions of $l_0$ 
as shown in Eq. (\ref{step_density}), which are to be determined by
Eq. (\ref{KM_of_step}).
The solutions of Eq. (\ref{relation_P}) under the condition (\ref{initial_P})
are ($l\ge 1$)
\bea
\gamma_1(l,k) &=& \rho^{l-1} \left [ \gamma_1 ( 1,k) - { 1-\rho\over 2} ( l -1)
\right ],\nonumber\\
\theta_1(l,k) &=& \rho^{l-1}
\left [ \theta_1(1,k) + \left ( \theta_0 - \rho\theta_0 +
{ 1\over 2\rho} \right )
{l-1\over 2} + { 1 - \rho \over 8\rho} (l-1)(l-2) \right ],
\label{solution_P} \\
\mu_1 (l,k) &=& \rho^{l-1}
\left [ \mu_1 ( 1, k) + { 1 - \rho \over 2} ( \mu_0 -k-2) (l-1)
- { 1 - \rho \over 4} l (l-1) \right ]
.\nonumber 
\eea
Now we can obtain the continuum limit of $R_{n+k}^l$.
Using Eqs. (\ref{recursive1}), (\ref{cont_P}), and (\ref{solution_P}),
we find the recursion relations of parameters given in Eq. (\ref{cont_lim})
\bea
\gamma(l,k) &=&\rho \gamma(l-1,k+1),
\nonumber \\
\theta (l,k) &=& \rho \theta(l-1,k+1) + \theta_1(l,k) +{\theta_0 \over 2}\rho^{l-1}
\left [ l  - \rho( l-1)   \right ],
\label{relation_R}
\nonumber\\
\mu (l,k) &=& \rho \mu(l-1,k+1) +  \mu_1 (l,k) + {\rho^{l-1} \over 2}( 1 + k +\mu_0 ) 
\left [ l - \rho (l-1)  \right ],
\nonumber
\eea
with conditions
\bea
\gamma(1,k) &=& 0,\nonumber\\
\theta(1,k) &=& { 1\over 4} + \theta_0 ( 1- \rho),\\
\mu(1,k) &=& ( 1 - \rho) \left ( \mu_0 - {1\over 2} \right ).\nonumber
\eea
We solve Eq. (\ref{relation_R}) step by step.
It is trivial to find that $\gamma(l,k) = 0$. 
As shown in Appendix \ref{Proof_No_KPZ}, 
the vanishing of $\gamma$ is not a coincidence.
Since $\theta_1(l,k)$ has no $k$ dependence, 
we expect that $\theta(l,k)$ 
also has no $k$ dependence. We find
\bea
\theta(l) =  l \rho^{l-1} 
\left [  {1\over 4} + \theta_0 ( 1 - \rho)  
+ \left ( 2 \theta_0 -2 \rho \theta_0 +  {1\over 2\rho} \right )
{ l-1 \over 4} + { 1-\rho \over 24 \rho}  (l-1) (l-2) 
\right ],
\label{kpzterm}
\eea
where $k$ is removed from the argument of $\theta$ due to the independence.
$\mu(l,k)$ seems to have an explicit $k$ dependence, but
by inserting $\mu_1(l,k)$ directly, we find the independence of $\mu(l,k)$
on $k$. The result reads (we drop $k$ by the same reason as $\theta$)
\be
\mu(l) = (1-\rho) \rho^{l-1} {l (l+1) \over 4}\left (
2 \mu_0 - {l\over 3} - {2\over 3} \right ),
\label{ewterm}
\ee

By requiring $C_n^S =0$ at the stationary state,
we can determine $\rho$, $\theta_0$, and
$\mu_0$ as a function of $l_0$. To find $\rho$, $\theta_0$, and
$\mu_0$, we need the continuum limit of 
\begin{equation}
\prod_{k=0}^l b_{n+k} \doteq \rho^{l+1} - {1\over 2} \rho^l ( l +1 ) D +
\rho^{l-1}{l+1 \over 8} ( l + 4 \rho \theta_0 ) D^2 +
\rho^l {l + 1 \over 4} ( 2 \mu_0 - l ) \partial D.
\end{equation}
We determine $\rho(l_0)$ from Eq. (\ref{KM_of_step}).
$\rho$ is the solution of the equation
\be
2 x^2 - 4 x + 1 - x^{2 l_0 + 1} \left [
1 - x ( 5 + 2 l_0 ) + x^2 ( 3 + 2 l_0) \right] = 0,
\label{MF_steady}
\ee
whose approximate solution is
\be
\rho (l_0 ) \simeq \rho_\infty \left ( 1 + { 1\over 4} \rho_\infty^{ 2 l_0 + 1}
( 2 l_0 + 1)\right ) + O(\rho_\infty^{4l_0}) ,
\ee
with $\rho_\infty = ( 2 - \sqrt{2})/2$, 
which is the solution for infinite $l_0$.
By the same reasoning, we can find $\theta_0$ and $\mu_0$, whose approximate
solutions read
\bea
\theta_0(l_0) &\simeq& { \sqrt{2} \over 4}
\left [ 1 + \rho_\infty^{ 2 l_0 }
\left ( {6 - 5 \sqrt{2} \over 4} + {21- 17 \sqrt{2} \over 6} l_0 + 
{2 -  \sqrt{2}\over 2} l_0^2 + {\sqrt{2}\over 3} l_0^3 \right )\right ]
+ O(\rho_\infty^{4l_0}),
\nonumber
\\
\mu_0 (l_0)&\simeq& { 4 - 3 \sqrt{2} \over 4}
\left [ 1 + \rho_\infty^{ 2 l_0 }
\left ( -{3\sqrt{2}\over 4}+
{4 - 17 \sqrt{2} \over 12} l_0 + 
{2 +  \sqrt{2}\over 2} l_0^2 + {2+ 2 \sqrt{2}\over 3} l_0^3 \right )
\right ] + O(\rho_\infty^{4l_0}).
\eea

For finite $l_0$, the most relevant terms arise from the second line
of Eq. (\ref{KM_of_height}) and the resulting equation is the KPZ one, with 
coefficients
\begin{eqnarray}
\nu = - \mu(2 l_0 + 1),\quad \lambda = - 2 \theta(2 l_0 + 1).
\end{eqnarray}
We give the numerical values of $\rho$, $\nu$, and $\lambda$ in the table.
These numbers are determined from the 
direct calculation of $\rho$, $\theta_0$ and $\mu_0$ using Eq. 
(\ref{KM_of_height}).
These $\lambda$'s should be compared with the previous 
numerical results\cite{KY97}. Note that
$\lambda$ in Ref.\cite{KY97} is one half 
of $\lambda$ here.

For infinite $l_0$, the second line of Eq. (\ref{KM_of_height}) vanishes,
of which the physical meaning is that 
a dropped particle eventually find a stable site. Hence  
the continuum equation becomes the VLD equation with coefficients
\begin{eqnarray}
\tilde \nu = - {1\over 2}\sum_{j=1}^\infty j \mu(j),\quad
\tilde \lambda = {1\over 2}\sum_{j=1}^\infty j \theta(j).
\end{eqnarray}
\section{~disappearance of odd powers of $D$ in $R$}
\label{Proof_No_KPZ}
This appendix proves that the first line of Eq. (\ref{KM_of_height}) does
not generate terms $\partial^2 (\partial h)^{ 2 i +1}$, where
$i$ is a non-negative integer.
To this end, consider a Taylor expansion of $R_{n+k}^l$ and set the
lattice constant to be $0$. 
We call this quantity $T_l^{l_0} (D)$,
and the corresponding quantity for $S$ $S_{l_0}(D)$.
One should not confuse this procedure with the continuum limit.
It is enough to prove that $T_l^{l_0} (-D) = T_l^{l_0} (D)$.
It is clear that $T_l^{l_0} (D)$ has no $k$ dependence because the 
lattice constant is set to $0$.
This fact enables us to write the recursion relation of $T_l^{l_0} (D)$
in the symmetric form
\bea
2 T_l^{l_0} ( D ) =  S_{l_0}(D) T_{l-1}^{l_0} (D) 
&& + \left ( 1 - {S_{l_0}(D)+D \over 2} \right ) \left (
{S_{l_0}(D) - D \over 2} \right )^ l \nonumber\\
&&+ \left ( 1 - {S_{l_0}(D)-D \over 2} \right ) \left (
{S_{l_0}(D) + D \over 2} \right )^ l.
\eea
Hence we see  that
\be
2 (T_l^{l_0} ( D ) - T_l^{l_0} ( -D )) =
S_{l_0}(D) ( T_{l-1}^{l_0} (D) - T_{l-1}^{l_0} (-D) )
\ee
The mirror symmetry of this system restricts the form of $S_{l_0}$ to be even 
in $D$,
so $T_{l-1}^{l_0} (D) = T_{l-1}^{l_0} (-D)$ implies
$T_l^{l_0} ( D ) =  T_l^{l_0} ( -D )$.
Indeed $T_1^{l_0} (D) = S_{l_0} + ( S_{l_0}^2 - D^2 ) /4$ and 
the logic of the mathematical induction 
proves the disappearance of the odd powers
of $D$ in $T_l^{l_0}(D)$.
The vanishing $\gamma(l,k)$ in appendix \ref{Rj} is a consequence of this
property.
As a result, we can safely affirm that CRSOS model is described
by the VLD equation.

\begin{table}
\caption{Numerical values of $\rho(l_0)$, $\nu(l_0)$, and $\lambda(l_0)$.}
\begin{tabular}{lccc}
$l_0$   &$\rho(l_0)$     &$\nu(l_0)$       &$\lambda(l_0)$\\\hline
0     &${1\over 3}$   &${1\over 3}$        &$-{5\over 6}$  \\
1     &0.299~027~750~50&$3.3887\times 10^{-1}$&$-9.5709\times 10^{-1}$\\
2     &0.293~696~759~81&$9.7310\times 10^{-2}$&$-2.9204\times 10^{-1}$\\
3     &0.292~988~255~45&$1.9569\times 10^{-2}$&$-6.0527\times 10^{-2}$\\
4     &0.292~903~676~47&$3.2656\times 10^{-3}$&$-1.0287\times 10^{-2}$\\
5     &0.292~894~314~95&$4.8299\times 10^{-4}$&$-1.5406\times 10^{-3}$\\
6     &0.292~893~329~94&$6.5674\times 10^{-5}$&$-2.1140\times 10^{-4}$\\
7     &0.292~893~229~81&$8.3964\times 10^{-6}$&$-2.7217\times 10^{-5}$\\
8     &0.292~893~219~88&$1.0242\times 10^{-6}$&$-3.3380\times 10^{-6}$\\
9     &0.292~893~218~92&$1.2038\times 10^{-7}$&$-3.9408\times 10^{-7}$\\
10    &0.292~893~218~82&$1.3730\times 10^{-8}$&$-4.5115\times 10^{-8}$\\
$\vdots$&$\vdots$&$\vdots$&$\vdots$\\
$\infty$&0.292~893~218~81&0&0\\
\end{tabular}
\end{table}
\begin{figure}
\caption{Relation of species to the height slope.
We use the integer to indicate the location of particles and
the half-integer for the height configuration. Hence the 
RSOS condition  at site $n+{1\over2}$ is determined by the situation
at sites $n$ and $n+1$.}
\label{config}
\end{figure}
\begin{figure}
\caption{Plots for the saturation width as a function of the system size
for various $l_0$. We find the saturation width by a least-squares fit
and the error bars represent three times the standard deviation, which includes
99\% of data. We fit the data as a function of $L$ and find two 
exponents. The exponent for the smaller system sizes are written in 
the bottom left and that of the larger system sizes in the upper right. 
The lines show
the fitting results. 
Up to 512, it seems plausible to insist that the systems are
in the scaling regime, but the data for $1024$ shows the clear
discrepancy within an error bar.
}
\label{alpha}
\end{figure}
\begin{figure}
\caption{Scaling plot of saturated widths of the one-dimensional 
RSOS/H model. The scaling variable is $l_0^\gamma / L$. 
The values of $l_0$ are equal to those in Fig. \ref{alpha}.
We also draw the line $x^{{\alpha_{\rm vld}}-{\alpha_{\rm kpz}}} = x^{0.4}$
as a guide to the eye.}
\label{1dcross}
\end{figure}
\newpage
\pagestyle{empty}
\centerline{\epsfysize=\textheight \epsfbox{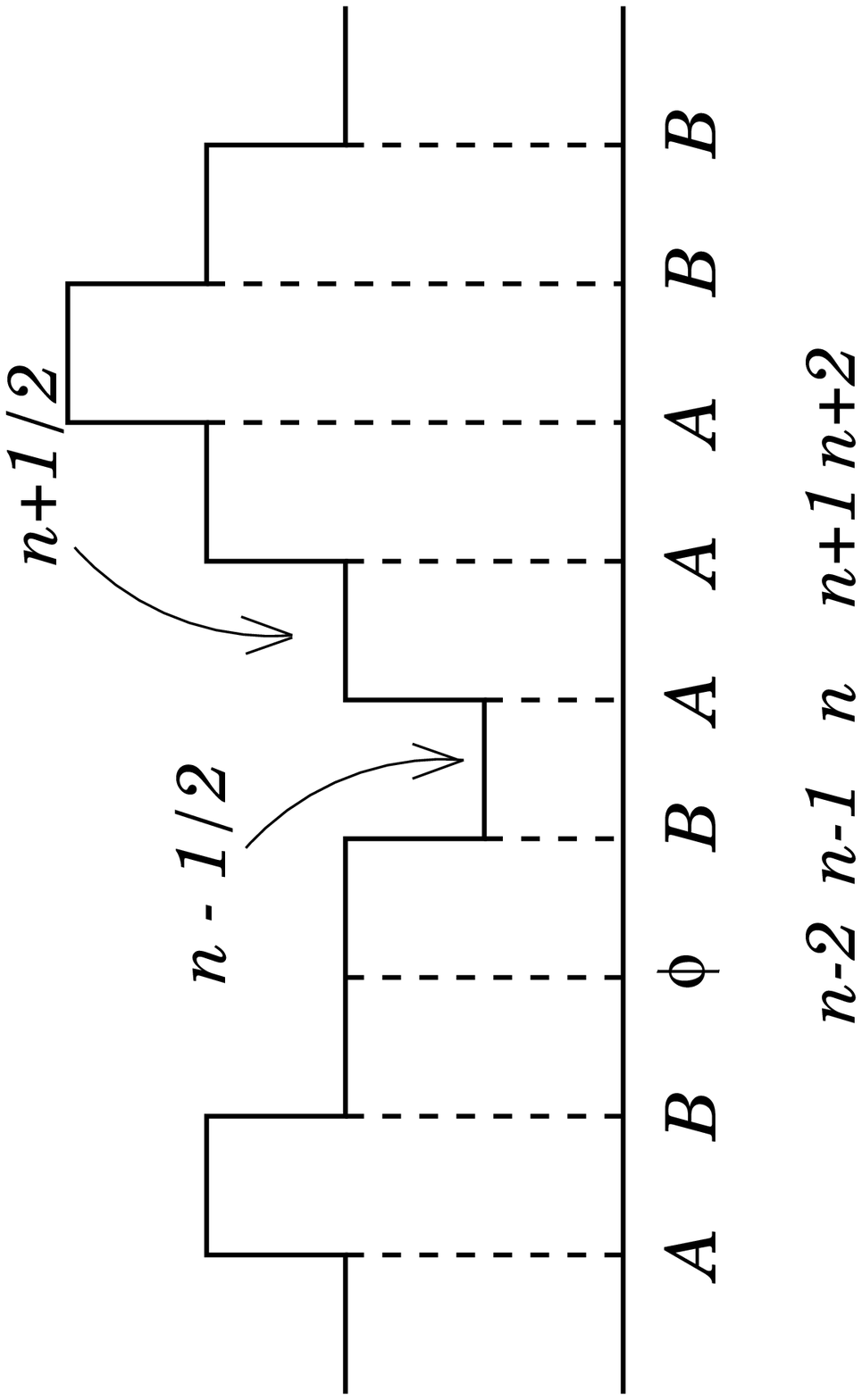}}
\centerline{\epsfysize=\textheight \epsfbox{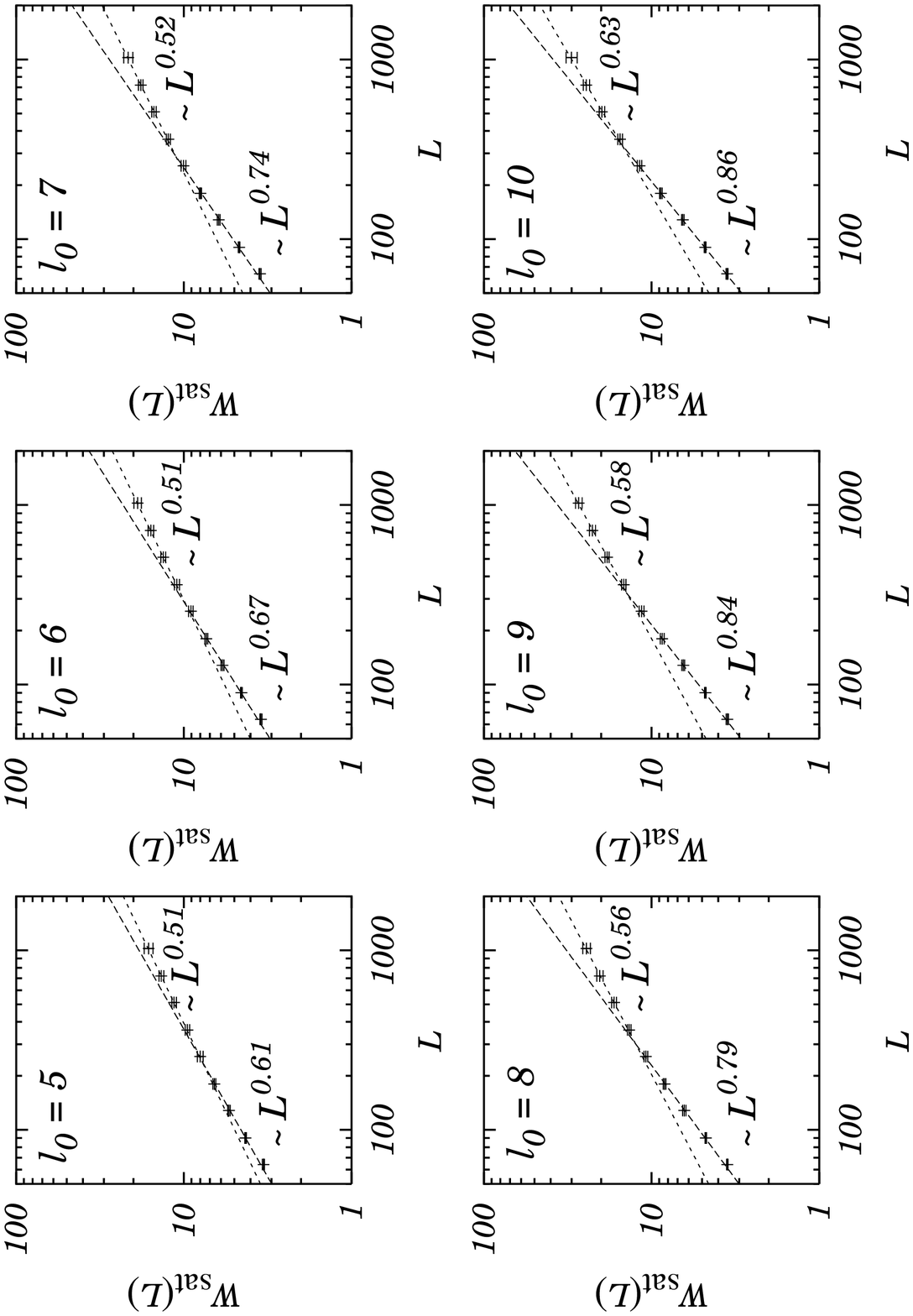}}
\centerline{\epsfysize=\textheight \epsfbox{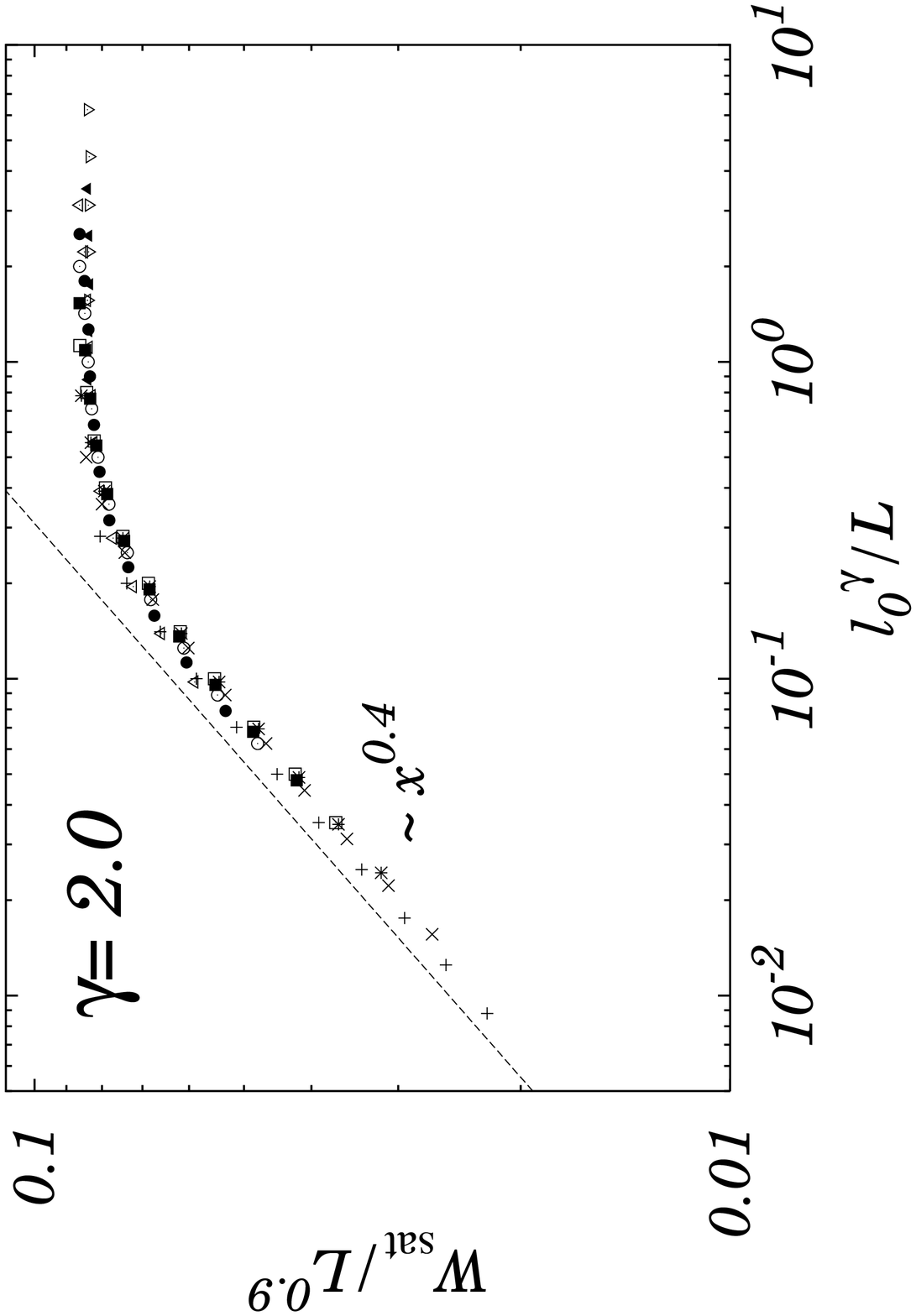}}
\end{document}